\documentclass[cameraready]{Interspeech}
\usepackage[table]{xcolor}  
\definecolor{Ablue}{HTML}{CFE3FF}
\definecolor{Bred}{HTML}{FFD6D6}

\title{How Attention Shapes Emotion: A Comparative Study of Attention Mechanisms for Speech Emotion Recognition}

\author[affiliation={1, 2}, orcid=0009-0003-9099-3826, correspondingauthor]{Marc}{Casals-Salvador}
\author[affiliation={1,2}, orcid=0000-0002-1389-3595]{Federico}{Costa}
\author[affiliation={1}, orcid=0000-0003-0192-7740, ]{Rodolfo}{Zevallos}
\author[affiliation={1, 2}, orcid=0000-0002-1730-8154, ]{Javier}{Hernando}

\address{
    $^1$ Barcelona Supercomputing Center, Spain \\
    $^2$ Universitat Politècnica de Catalunya, Spain
}

\email{marc.casals@bsc.es, rodolfo.zevallos@bsc.es, federico.costa@bsc.es, javier.hernando@bsc.es}

\usepackage{comment}

\begin{document}

\maketitle

\begin{abstract}
    Speech Emotion Recognition (SER) plays a key role in advancing human–computer interaction. Attention mechanisms have become the dominant approach for modeling emotional speech due to their ability to capture long-range dependencies and emphasize salient information. However, standard self-attention suffers from quadratic computational and memory complexity, limiting its scalability. In this work, we present a systematic benchmark of optimized attention mechanisms for SER, including RetNet, LightNet, GSA, FoX, and KDA. Experiments on both MSP-Podcast benchmark versions show that while standard self-attention achieves the strongest recognition performance across test sets, efficient attention variants dramatically improve scalability, reducing inference latency and memory usage by up to an order of magnitude. These results highlight a critical trade-off between accuracy and efficiency, providing practical insights for designing scalable SER systems.
\end{abstract}

\section{Introduction}

Speech Emotion Recognition (SER) is a fundamental component of interactive computing, enabling the automatic inference of human emotional states from speech signals. To foster progress in this area, several benchmark datasets have been established, including IEMOCAP \cite{busso2008iemocap}, RAVDESS \cite{livingstone2018ryerson}, MELD \cite{poria2019meld}, and MSP-Podcast \cite{Busso_2025,8003425}. Within these benchmarks, attention-based models \cite{li2019improved} have emerged as the dominant paradigm due to their ability to capture long-range temporal dependencies and contextual relationships in emotional speech.

Despite their strong performance, softmax attention (SA) mechanisms suffer from quadratic computational and memory complexity with respect to sequence length. This limitation is particularly relevant in SER, where speech signals may span long durations and exhibit high temporal redundancy. As a result, full SA introduces substantial computational overhead, which can hinder scalability and limit applicability in resource-constrained environments.

To address these limitations, several efficient attention mechanisms have been proposed to reduce computational cost while preserving modeling capacity. These approaches introduce alternative mechanisms for information aggregation, such as retention-based memory, gated accumulation, and forgetting-based dynamics. Architectures including RetNet \cite{Sun2023RetentiveNA}, LightNet \cite{qin2024scanonceefficientmultidimension}, Gated Slot Attention (GSA) \cite{zhang2024gsa}, Forgetting Transformer (FoX) \cite{lin2025forgetting}, and Kimi Delta Attention (KDA) \cite{zhang2025kda} aim to improve scalability by reducing memory usage and computational complexity while maintaining competitive performance.

However, existing SER research primarily focuses on recognition accuracy, often overlooking efficiency-related factors such as memory consumption, training time, and inference speed. This omission limits the practical applicability of SER systems, particularly in real-world scenarios where computational resources are constrained. Furthermore, to our knowledge, no prior work has systematically benchmarked modern efficient attention architectures for SER under a unified experimental setting that jointly evaluates recognition performance and computational efficiency.

In this paper, we present a systematic analysis of efficient attention architectures for SER. Specifically, we evaluate RetNet, LightNet, GSA, FoX, and KDA on the MSP-Podcast corpus \cite{Busso_2025,8003425}, comparing their performance in terms of emotion recognition accuracy, training memory consumption, and inference efficiency. Our study provides a comprehensive analysis of the trade-offs between recognition performance and computational efficiency, offering practical insights into the suitability of these architectures for scalable and resource-efficient SER systems.

The remainder of this paper is organized as follows. Section \ref{sec:related-work} briefly describes the use of attention-based mechanisms in SER, from merging different modalities to the usage of transformer-based encoders. Section \ref{sec3} introduces our architecture and deeply explains its design.  It also introduces the different seq2seq alternatives.  Section \ref{sec4} presents the experimental setup, including dataset splits, data preprocessing, backbone encoders, and the evaluation metrics. Section \ref{sec5} presents quantitative results in two versions of MSP-Podcast, highlighting the trade-offs between accuracy and computational cost. Finally, Section \ref{sec6} concludes the paper with a discussion that remarks practical implications and provides an interpretation of the results.

\section{Related Work}
\label{sec:related-work}

\begin{figure*}[h]
    \centering
        \includegraphics[width=\linewidth]{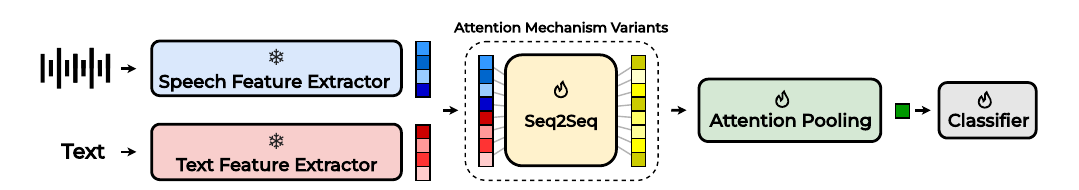}
    \caption{System's architecture. Experiments are made considering different attention mechanisms for the seq2seq module.}
    \label{fig:system_diagram}
    \vspace{-0.3cm}
\end{figure*}

Attention mechanisms have been used in SER to address the variable length and non-uniform emotional salience of speech signals.
Early studies demonstrated that local attention over Recurrent Neural Networks' hidden states significantly improved performance by emphasizing emotionally informative regions of speech \cite{mirsamadi2017automatic}.
Subsequent research expanded attention usage through self-attention mechanisms to focus on the salient periods of emotion in speech utterances \cite{li19n_interspeech}.

In recent years, a series of works have leveraged self-attention and transformer-based encoders to capture complex emotional patterns within acoustic and linguistic modalities. 
The Odyssey 2024 Speech Emotion Recognition Challenge report shows that most multimodal competitive systems extract speech and text representations using self-supervised pre-trained transformer-based models, and fuse them using statistic or attentive poolings \cite{goncalves24_odyssey, chen24_odyssey, lafore24_odyssey, shamsi24_odyssey}.
A double multi-head attention mechanism is proposed in \cite{costa24_odyssey} as an early fusion strategy, where a multi-head attention layer transforms these mixed features into complementary contextualized representations, followed by a second attention mechanism applied to pool these representations into an utterance-level vector.
In \cite{harm24_odyssey}, a convolution-augmented transformer \cite{gulati20_interspeech} speech feature extractor is used. 

The Interspeech 2025 challenge on speech emotion recognition in naturalistic conditions \cite{naini2025interspeech} also highlights the widespread use of self-attention and transformer encoders across submitted systems \cite{chatzichristodoulou25_interspeech, shi2025advancing}.
In \cite{ueda25_interspeech}, several cross-modal models are obtained using combinations of pre-trained features from different modalities, and trained independently with a fusion mechanism using cross-modal attention.
This mechanism consists of a multi-head attention layer that takes one modality as the query and another modality as the key and value.

SER research extensively validates the effectiveness of classic attention mechanisms, including local attention, self-attention, multi-head attention, and attentive pooling, across unimodal and multimodal settings. 
While these works demonstrate the maturity of classic attention in SER, they also reveal a clear gap: efficient attention mechanisms designed to reduce computational overhead have not been systematically evaluated for emotional speech modeling.
This motivates the present work, which systematically benchmarks modern efficient attention mechanisms against classic self-attention to assess their suitability for the unique temporal and acoustic characteristics of emotional speech.

\begin{table*}[h]
\centering
\fontsize{7}{8}\selectfont
\setlength{\tabcolsep}{6pt}
\renewcommand{\arraystretch}{1.15}

\begin{tabular}{l|ccc|ccc|ccc|ccc|ccc}
\toprule
& \multicolumn{3}{c|}{\textbf{Wav2Vec2XLSR}}
& \multicolumn{3}{c|}{\textbf{WavLM}}
& \multicolumn{3}{c|}{\textbf{Wav2Vec2}}
& \multicolumn{3}{c|}{\textbf{HuBERT}}
& \multicolumn{3}{c}{\textbf{Mean}} \\
\textbf{Mechanism}
& \textbf{Dev} & \textbf{T1} & \textbf{T2}
& \textbf{Dev} & \textbf{T1} & \textbf{T2}
& \textbf{Dev} & \textbf{T1} & \textbf{T2}
& \textbf{Dev} & \textbf{T1} & \textbf{T2}
& \textbf{Dev} & \textbf{T1} & \textbf{T2} \\
\midrule
\textbf{Baseline}
& - & -& -
& - & 29.7 & 20.6
& - & 23.8 & 15.6
& - & 28.5 & 19.2
& -  & 27.3 & 18.5 \\
\midrule
\textbf{SA}
& 37.04 & \textbf{36.63} & \textbf{27.34}
& 37.50 & \textbf{36.80} & \textbf{27.30}
& 34.83 & \textbf{34.95} & \textbf{26.06}
& 36.18 & \textbf{37.31} & \textbf{28.07}
& 36.39 & \textbf{36.42} & \textbf{27.19} \\

\textbf{RetNet}
& 33.70 & 33.40 & 24.18
& 34.93 & 34.74 & 24.85
& 36.05 & 34.59 & 24.48
& 34.99 & 34.46 & 24.86
& 34.92 & 34.30 & 24.59 \\

\textbf{LightNet}
& \textbf{38.11} & 34.59 & 25.56
& \textbf{37.73} & 33.54 & 25.19
& \textbf{36.23} & 33.97 & 25.10
& 34.40 & 33.63 & 24.90
& \textbf{36.62} & 33.93 & 25.19 \\

\textbf{GSA}
& 25.74 & 26.32 & 18.61
& 36.94 & 29.66 & 19.53
& 35.77 & 31.96 & 23.80
& 32.48 & 33.01 & 24.97
& 32.73 & 30.24 & 21.73 \\

\textbf{FoX}
& 35.35 & 34.44 & 25.12
& 36.30 & 34.65 & 23.77
& 35.65 & 34.82 & 25.39
& \textbf{36.22} & 35.88 & 26.96
& 35.88 & 34.95 & 25.31 \\

\textbf{KDA}
& 34.13 & 34.50 & 26.88
& 31.50 & 32.95 & 23.13
& 34.09 & 33.70 & 25.11
& 34.26 & 30.72 & 22.73
& 33.50 & 32.97 & 24.46 \\

\bottomrule
\end{tabular}

\caption{Macro F-score (\%) for different attention mechanisms on MSP-Podcast. Dev results use MSP-Podcast v1.0 \cite{8003425}, while T1 (Test1) and T2 (Test2) use MSP-Podcast v2.0 \cite{Busso_2025}. Mean averages across SSL backbones.}
\label{table:f-scores}
\vspace{-0.3cm}
\end{table*}

\section{Methodology}
\label{sec3}
\subsection{Overview}

This work investigates the impact of the different sequence-to-sequence (seq2seq) attention mechanisms on performance and efficiency in the speech emotion recognition task (SER). We focus on the fusion stage, where speech and text representations are merged into one single sequence capable of modeling emotional features. To isolate the effect of these different mechanisms, all other architectural components, including the distinct feature extractors, the pooling strategy, the classifier, and the training protocol, are fixed across the experiments. By benchmarking multiple seq2seq attention alternatives, this study provides empirical evidence of how attention design affects emotional representation learning. Figure  \ref{fig:system_diagram} represents the system used for the experiments. We provide the code\footnote{To ensure author anonymity, the link to the resource will be added after the review process} to ensure reproducibility.

\subsection{Proposed Architecture}
Let $X=[x_1,...,x_T]$ and $W=[w_1,...,w_N]$ be a raw audio waveform and a tokenized text. The model computes an 8-class prediction $\hat{y}\in\{1,...,8\}$. Those modalities are processed by speech and text feature extractors.

\begin{align}
    S&=f_{\text{speech}}(X)\in\mathbb{R}^{M \times D} \\
    E&=f_{\text{text}}(W)\in\mathbb{R}^{N\times D}
\end{align}

Where $M$ and $N$ correspond to the sequence lengths produced by the feature extractors and $D$ is their hidden size.
The embeddings are concatenated $U = [S;E]\in\mathbb{R}^{L\times D}$ where $L = M + N$. The seq2seq is a function $\mathcal{M}_\theta: \mathbb{R}^{L\times D}\rightarrow\mathbb{R}^{L\times D}$, hence we define the hidden vector as $H=\mathcal{M}_\theta (U)$. This vector is processed by attention pooling to obtain a global representation $c$:

\begin{align}
    \omega_i &= \frac{\exp\left(\frac{h_i^\top q}{\sqrt{D}}\right)}{\sum_{l=1}^{L}\exp\left(\frac{h_l^\top q}{\sqrt{D}}\right)}\\ 
    c &= \sum_{l=1}^{L}\omega_l h_l \in\mathbb{R}^D
\end{align}
where $q$ is a trainable parameter.

The final unit $c$ is processed by a classifier which encompasses dropout, GELU activation function, and layer normalization.

\subsection{Sequence-to-sequence alternatives}

Standard self-attention is not the only paradigm for seq2seq modeling, and several recent approaches have been proposed to improve scalability and inductive bias for long sequences. In this subsection, we briefly review representative alternatives that replace or modify self-attention while retaining strong contextual modeling capabilities.

\begin{itemize}

\item \textbf{Softmax Attention \cite{NIPS2017_3f5ee243}.}  
The transformer self-attention mechanism explicitly models pairwise interactions between all elements in a sequence using content-based similarity. While highly expressive, it exhibits quadratic time and memory complexity with respect to sequence length, which limits its applicability to long or high-resolution inputs.

\item \textbf{RetNet \cite{Sun2023RetentiveNA}.}  
RetNet replaces softmax-based attention with a retention mechanism that accumulates past information using exponential decay. By introducing a controllable forgetting factor, RetNet balances long-term memory with recency bias. Its formulation supports both parallel and recurrent computation, enabling efficient modeling of long sequences without explicitly constructing attention matrices.

\item \textbf{LightNet \cite{qin2024scanonceefficientmultidimension}.}  
LightNet proposes a linear-time sequence modeling approach based on additive recurrence. Instead of relying on multiplicative decay or attention weights, LightNet aggregates contextual information through a single-pass scan with an additive decay parameter. This design avoids attention altogether while preserving the ability to capture long-range dependencies efficiently.

\item \textbf{Gated Slot Attention (GSA) \cite{zhang2024gsa}.}  
GSA combines slot-based representations with gating mechanisms to regulate information flow over time. By maintaining a fixed number of latent slots and selectively updating them, GSA constrains memory growth while enabling structured and persistent representations, which is especially beneficial for long sequences inputs.

\item \textbf{Forgetting Transformer (FoX) \cite{lin2025forgetting}.}  
FoX augments standard attention with an explicit forgetting mechanism that progressively downweights older tokens. Rather than treating all past context equally, FoX introduces a learnable temporal decay that biases the model toward more recent information, improving robustness for long sequences while retaining the Transformer structure.

\item \textbf{Kimi Delta Attention (KDA) \cite{zhang2025kda}.}  
KDA extends forgetting-based attention by introducing fine-grained, dimension-wise decay. Instead of applying a uniform forgetting factor, KDA dynamically controls information retention at the feature level, allowing different representation dimensions to decay at different rates. This results in a more expressive and adaptive mechanism for long-context modeling.

\end{itemize}

\section{Experimental Setup}
\label{sec4}
\subsection{Dataset}
The experiments were conducted using MSP-Podcast versions 1 \cite{8003425} and 2 \cite{Busso_2025}. This corpus comprises spontaneous speech segments from various audio-sharing websites. It includes eight categorical emotions: anger, happiness, sadness, fear, surprise, contempt, disgust, and a neutral state, as well as attribute-based annotations of valence, arousal, and dominance.

The first version of the dataset is composed of two splits: training and development (Dev). In contrast, the newest version expands this corpus by incorporating two test sets with available labels: Test1, created using the retrieval-based protocol and reflecting the corpus’s natural emotion distribution; Test2, a control split generated without emotion-based retrieval to assess potential selection bias.

\subsection{Feature Extraction}
For this work, we have used and tested multiple pretrained self-supervised learning (SSL) models to enhance the extraction of emotional attributes. For both modalities, feature extractors use large versions; hence, $D=1024$.

\textbf{Speech Features Extractor:} We employed transformer-based models, including WavLM \cite{9814838}, Wav2Vec2 \cite{baevski2020wav2vec}, HuBERT \cite{10.1109/TASLP.2021.3122291}, and Wav2Vec2XLSR \cite{conneau21_interspeech}. All these models process raw audio waveforms directly.

\textbf{Text Features Extractor:} To extract linguistic information from the text, we have used BERT-like models such as BERT \cite{devlin-etal-2019-bert}, RoBERTa \cite{liu2019robertarobustlyoptimizedbert}, and Modern BERT \cite{warner-etal-2025-smarter}. In preliminary experiments, BERT large uncased was the feature extractor that achieved the best F-score. This model comprises 340M parameters and 24 Transformer layers and is pre-trained on BookCorpus \cite{10.1109/ICCV.2015.11} and English Wikipedia.

\subsection{Training Details}

The audio waveforms are normalized using the mean and standard deviation of the training set. They are then cropped into segments of 5.5 seconds, which gave the best validation Macro F-score in our preliminary tuning. In this study, no data augmentation techniques were employed. The modality feature extractors are frozen during the training, and the only learnable parameters are in the seq2seq, the Attention Pooling, and the classifier.

All models are trained for 20 epochs, with checkpoints saved whenever the model surpasses the validation F-Score. The batch size selected is 32. The model parameters are optimized using AdamW \cite{Loshchilov2017DecoupledWD} with a learning rate of \texttt{1e-4} that is halved every 5 epochs without a validation improvement. Regarding the seq2seq mechanism, four heads were used with a dropout of 0.4. To ensure fair comparison across attention mechanisms, all models are configured to have approximately the same number of trainable parameters 20M, while the total parameter count is $\approx$655M due to frozen pre-trained encoders.

All experiments have been performed using the PyTorch framework \cite{NEURIPS2019_bdbca288} in conjunction with Flash Linear Attention \cite{yang2024fla} using four NVIDIA H100. All models were trained under identical hardware constraints.
\subsection{Evaluation Details}
The model's evaluation comprises two distinct components: performance and efficiency.

In the first case, the model is evaluated using the Dev partition of the MSP-Podcast, as well as the Test1 (T1) and Test2 (T2) sets of MSP-Podcast 2.0. Waveforms are normalized using the mean and standard deviation computed from the training data. The whole audio file is used, which implies that the batch size to evaluate is equal to one. The metric used for this evaluation is the Macro F-Score, which is particularly useful for addressing the class imbalance presented. Different SSLs are tested to benchmark all configurations. The baseline given by the dataset’s paper consists of different speech feature extractors with a simple 2-layer FC head that learns emotion classification and valence/arousal/dominance regression simultaneously.

In the second case, the efficiency of the seq2seq layer is analyzed by comparing two metrics: the inference time (the mean time it takes to process a sample), and the Peak GPU usage (the GPU allocation required to perform the mechanism). In both scenarios, a batch size of one is used while varying the sequence length. Random cropping is applied when sequences are shorter than the target length, and repetition padding is used when sequences need to be longer. Given that the seq2seq layer represents a limited portion of the network, changes in latency and memory can be overshadowed by the other parts of the architecture. To highlight the differences in this module, all efficiency measurements are conducted exclusively on the seq2seq module, with the speech feature extractor fixed as Wav2vec2XLSR.

\begin{figure}[t]
    \centering
        \includegraphics[width=1\linewidth]{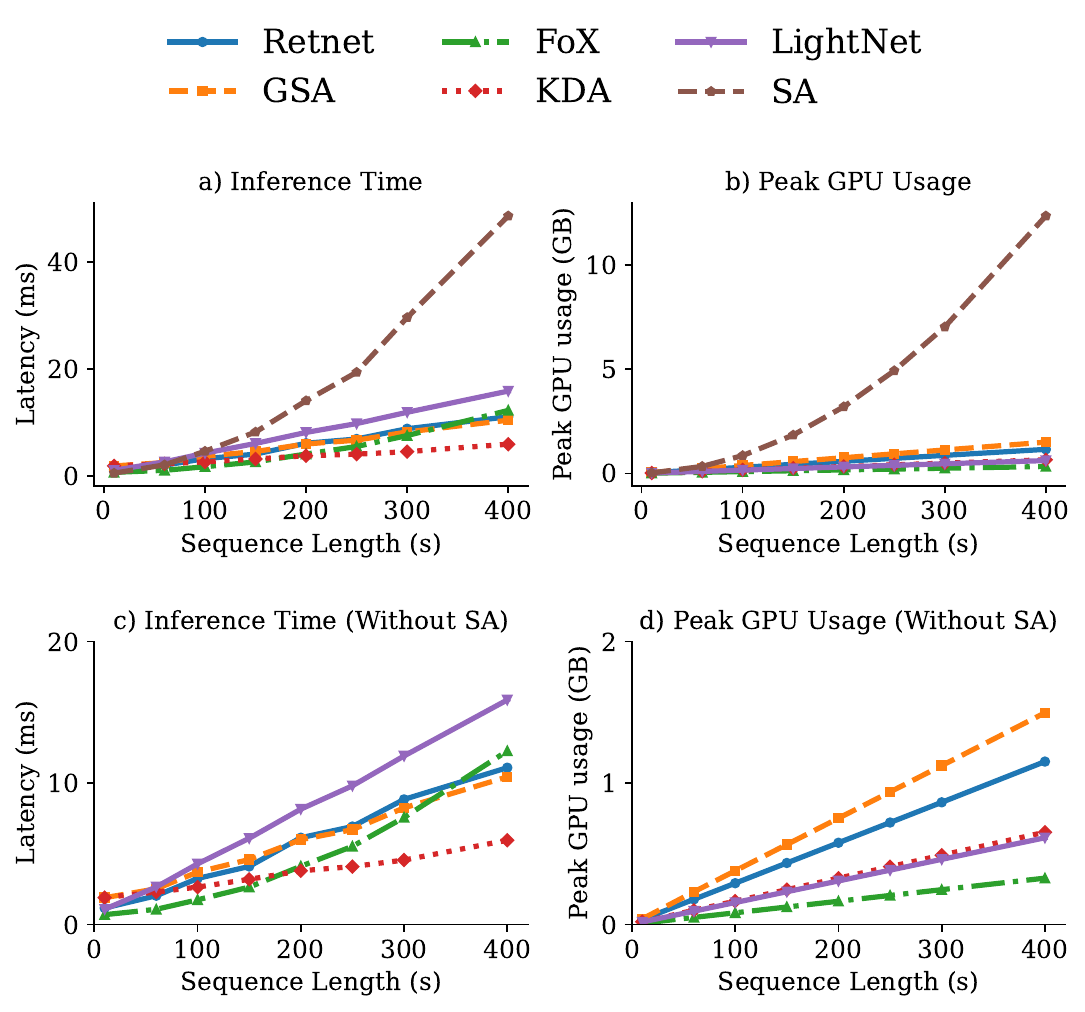}
\caption{Inference time and peak GPU memory usage of the seq2seq module as a function of sequence length on the MSP-Podcast dev set \cite{8003425}. Panels (a–b) report results for all models. Panels (c–d) provide a zoomed view excluding SA to make the relative growth trends of the remaining alternatives easier to distinguish.}
\label{fig:inference-time-memory}
\end{figure}
\section{Experimental Results}
\label{sec5}
Table \ref{table:f-scores} summarizes the Macro F-score on MSP-Podcast (Dev, T1, T2) for four SSL backbones. On the development set, LightNet attains the best mean performance (36.62\%), narrowly ahead of SA (36.39\%), and yields the top Dev score for three backbones, including the strongest single result (38.11\% on Wav2Vec2XLSR). On the evaluation splits, SA generalizes best, achieving the highest mean on both T1 (36.42\%) and T2 (27.19\%) and leading across all backbones on T1. FoX is consistently second (34.95\% on T1; 25.31\% on T2), while RetNet and LightNet drop relative to Dev. As expected, all methods degrade on T2, with GSA showing the largest instability and the weakest robustness (21.73\% mean on T2). Overall, although LightNet is strongest on Dev, SA provides the most stable and best-performing results on the evaluation splits.

Figure \ref{fig:inference-time-memory} reports seq2seq-module inference latency (ms) and peak GPU memory (GB) as a function of sequence length on the MSP-Podcast Dev set (batch size = 1). Efficient alternatives exhibit approximately linear scaling in both compute and memory, whereas SA grows quadratically with sequence length. Consequently, SA remains competitive for short inputs (0.55 ms at 10 s) but becomes substantially more expensive for long sequences (48.59 ms at 400 s). In contrast, KDA requires only 5.96 ms at 400 s, corresponding to an 8.15$\times$ latency reduction. Memory follows the same trend: SA increases from 0.02 GB to 12.35 GB, while all efficient variants remain below 1.50 GB; FoX is the most memory-efficient, using 0.328 GB at 400 s (37.6$\times$ less than SA). Overall, efficient seq2seq operators scale markedly better for long sequences, offering clear speed–memory trade-offs (KDA fastest, FoX most memory-efficient) while avoiding SA’s quadratic growth.

\section{Discussion}
\label{sec6}

The benchmark highlights a consistent trade-off between recognition performance and computational efficiency in SER by systematically comparing SA and several recent linear or recurrent alternatives within a unified architecture. The study offers valuable insight into the trade-offs between modeling capacity, inference latency, and memory efficiency for long-duration audio understanding. The results demonstrate that modern efficient architectures can approach the performance of standard SA while substantially improving runtime and memory characteristics.

The models explored can be broadly categorized into unbounded memory and bounded memory models. The first category creates an explicit representation of past speech representations, leading to memory requirements that grow with sequence length. Within this category, it is possible to find SA and FoX. On the contrary, bounded memory models compress past information into a fixed-dimensional recurrent state.

Let $L$ be the sequence length and $D$ be a fixed dimension, the theoretical latency of unbounded models is $O(L^2D)$. This is clearly reflected in the SA results in Figure \ref{fig:inference-time-memory}. On the contrary, the FoX implementation achieves low latency due to kernel optimizations implicit in the implementation. In contrast, bounded models have a theoretical latency of $O(kL)$ where $k$ is a fixed value that does not depend on sequence length. As represented on the plot, these models outperform the vanilla SA.

From the memory point of view, SA still materializes the full attention matrix, making it $O(L^2)$. Fused kernels, such as those used by FoX, reduce the peak activation memory to $O(LD)$, being a more optimized option. Conversely, models with theoretically bounded memory (e.g., KDA, RetNet) exhibited higher peak memory usage than FoX in our full-sequence evaluation. 

A second important finding is that all methods experience a substantial drop in performance from Test1 to Test2.  In particular, Test2 appears to better reflect realistic audio conditions and class imbalance, resulting in a more noticeable degradation across all models. This indicates room for architectural improvements for this task.

In summary, our results show a clear accuracy–efficiency trade-off: softmax attention remains the strongest choice when peak performance on short inputs is required, whereas efficient mechanisms provide substantially better scaling for long sequences. Finally, the drop from Test1 to Test2 indicates that improving robustness under real-world audio scenarios is the key direction for closing the remaining performance gap.
\clearpage

\ifcameraready

\section{Acknowledgments}

\else
\fi

\section{Generative AI Use Disclosure}
The code presented in this article was developed with the assistance of GitHub Copilot programming aids. Additionally, ChatGPT and Grammarly were utilized to enhance spelling, grammar, clarity, and overall readability during manuscript preparation.

\bibliographystyle{IEEEtran}
\bibliography{mybib}

\end{document}